\documentclass[sigchi]{acmart} 
\renewcommand\footnotetextcopyrightpermission[1]{}
\AtBeginDocument{%
  }

\usepackage{subcaption}

\acmConference[CHI '25]{April 26 - May 01,
  2018}{Yokohama, JPN}

\begin{document}

\title{No Fuss, Just Function - A Proposal for Non-Intrusive Full Body Tracking in XR for Meaningful Spatial Interactions}


\author{Elisabeth Mayer}
\email{elisabeth.mayer@lrz.de}
\orcid{0000-0002-2551-0846}
\author{Thomas Odaker}
\email{thomas.odaker@lrz.de}
\orcid{0000-0003-0675-8675}
\affiliation{%
 \institution{Leibniz Supercomputing Centre}
  \city{Garching near Munich}
  \country{Germany}
}
\author{Dieter Kranzlmüller}
\email{dieter.kranzlmueller@lrz.de}
\orcid{0000-0002-8319-0123}
\affiliation{%
  \institution{LMU Munich}
  \city{Munich}
  \country{Germany}
}

\renewcommand{\shortauthors}{Mayer et al.}

\begin{abstract}
 Extended Reality (XR) is a rapidly growing field with a wide range of hardware from head mounted displays to installations. Users have the possibility to access the entire Mixed Reality (MR) continuum. Goal of the human-computer-interaction (HCI) community is to allow natural and intuitive interactions but in general interactions for XR often rely on handheld controllers. One natural interaction method is full body tracking (FBT), where a user can use their body to interact with the experience. Classically, FBT systems require markers or trackers on the users to capture motion. Recently, there have been approaches based on Human Pose Estimation (HPE), which highlight the potential of low-cost non-intrusive FBT for XR. Due to the lack of handheld devices, HPE may also improve accessibility with people struggling with traditional input methods. This paper proposes the concept of non-intrusive FBT for XR for all. The goal is to spark a discussion on advantages for users by using a non-intrusive FBT system for accessibility and user experience. 
\end{abstract}



\begin{teaserfigure}
  \includegraphics[width=\textwidth]{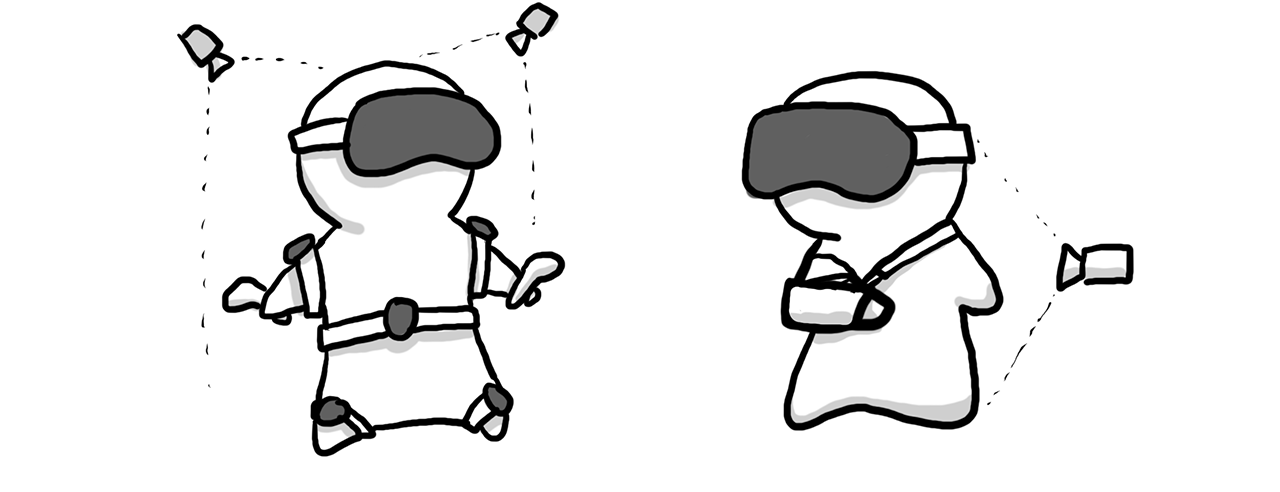}
  \caption{Intrusive Full Body Tracking vs Non-Intrusive Full Body Tracking for XR: One user wearing marker-based FBT (left) and one user with a temporary motor limitation using non-intrusive FBT (right).}
  \Description{Intrusive Full Body Tracking vs Non-Intrusive Full Body Tracking for XR: One user wearing marker-based FBT (left) and one user with a temporary motor limitation using non-intrusive FBT (right).}
  \label{fig:teaser}
\end{teaserfigure}


\maketitle

\section{Introduction}

Over the course of the last few years, Extended Reality (XR) has been a rapidly developing field \cite{lopes2023user, martin2017virtual}. A wide variety of devices ranging from head-mounted-displays (HMDs) \cite{Anthes2016} to large-scale installations like CAVE systems \cite{cruz1992cave} can be used to access the Mixed Reality (MR) continuum \cite{milgram1995augmented}. Devices like the Meta Quest 3\footnote{\href{https://www.meta.com/de/quest/quest-3/}{Meta Quest 3 Website (accessed last 28.02.2025)}} or Apple Vision Pro\footnote{\href{https://www.apple.com/apple-vision-pro/}{Apple Vision Pro Website (accessed last 28.02.2025)}} have started to blur the line between Augmented Reality (AR) and Virtual Reality (VR), providing new use-cases and easier accessibility to MR. 
One key aspect of MR has always been interaction and a primary goal of human-computer-interaction (HCI) is to allow for natural and intuitive interaction techniques so users can focus on the experience provided \cite{nielsen2004procedure}. 

Interaction in XR often relies on the use of input devices like handheld controllers. While this method is well established, it may not be intuitive for everyone and may exclude people with disabilities who struggle with the use or of such devices. Another way to provide natural interaction is full body tracking (FBT), providing data about a user's full body. In the past, FBT often required the use of body-mounted devices like markers or trackers. Recent advances in human pose estimation (HPE), however, seem a promising development to enable FBT without body-mounted devices or complex camera setups. HPE could provide a way to create a low-cost, non-intrusive tracking system for FBT \cite{capece2018low}. Due to the lack of handheld devices, HPE may also improve inclusivity and provide universal access to MR for people struggling with traditional input methods for various reasons. 

In this paper, we want to propose a concept for a non-intrusive full body tracking system based on HPE for the use with MR environments. The goal is to spark a discussion about accessibility in MR and move towards designing tracking systems and interactions suitable for everybody.

\section{Related Work}
This paper focuses on the spatial affordance \cite{vieira2024understanding}, specifically in terms of interactions and movements within a 3D space. For spatial interactions 3D User Interfaces (3UI) are involved. 3DUIs are inherently spatial with tasks ranging from selection to navigation that require users and their input devices to interact within a 3D space \cite{bowman2001introduction}. 

In general, XR can be experienced through HMDs, like the Meta Quest 3. Interactions for headsets usually can be activated through controllers \cite{10.1145/3613905.3650925, angelov2020modern}. These controllers are typically handheld and interactions are triggered through button presses \cite{Anthes2016}. However, these controllers can be unfamiliar for novice users and are dependent on users grip posture and finger movements which can be a challenge for some users \cite{dudley2023inclusive}. Some headsets now offer hand tracking or gesture control \cite{papadopoulos2023vrgestures}.

One natural way for users to interact in XR is through FBT, which enables tracking the position and orientation of a user's body for interactions \cite{theodoropoulos2023developing}. Input with FBT allows users to directly interact with the XR experience with their own body or body movement \cite{kulik2009building, yang2022hybridtrak}.
One advantage of FBT is that users do not have to learn a new form of input as it can map their natural movement to the experience \cite{steed2021directions}.  FBT can help users to feel more immersed in an experience as they can focus on the experience itself over the controllers and interactions that they are required to perform \cite{caserman2019survey}. Inman et al. found that by using FBT for learning how to drive a wheel chair in VR users praticed more and learnt faster \cite{inman2011learning}.

One form of FBT tracking for XR, e.g. AR \cite{jun2010extended} and VR, is marker-based systems \cite{caserman2019real}. While marker-based systems can be very accurate, they can be expensive and by requiring external trackers or markers, users might feel uncomfortable and require a prolonged setup time \cite{caserman2019survey}. In terms of accessibility in XR, any additional hardware can provide a challenge for people who need to wear hardware over eyeglasses or hearing aids \cite{mott2019accessible}.

Markerless FBT systems do not require the users to wear any external devices \cite{caserman2019real, comport2003real}. Markerless systems are generally affordable and only require a simple setup \cite{caserman2019survey}. Previously a term commonly used when describing eye-tracking \cite{kim2004non}, non-intrusive tracking describes a technology where users are not required to wear any devices, which marker-less systems offer. Non-intrusive tracking for FBT was identified as vital for HCI. Colombo et al. expand the term non-intrusive to encompass low-cost as an advantage of non-intrusive tracking \cite{956993}. Gabel et al. also iterate the importance of non-intrusive low-cost for movement analysis \cite{gabel2012full}. 

One form of low cost markerless FBT is monocular FBT by using a single input sensor \cite{caserman2019survey}. With this method, there are some detection and tracking issues based on occlusion and lighting \cite{sminchisescu2001covariance, helten2013full,caserman2019survey}. To solve these issues, a method from computer vision can be applied: Human Pose Estimation (HPE) \cite{desmarais2021review}. 
HPE is the estimation of the position of humans and their joints based on images, videos or depth sources \cite{o1980model}. HPE is a heavily researched field in computer vision \cite{lan2022vision} and can also be used for XR experiences \cite{jo2023enhancing}. HPEs offer a low-cost FBT solution and can be dynamically adapted for users with upper- and lower-body motor limitations \cite{huang2024wheelpose}.

\section{Non-Intrusive Full Body Tracking}
 One major affordance of XR is the spatial affordance, where the users move and interact within a 3D space. In the past, FBT typically required the use of additional devices. However, HPEs offer a promising solution for low-cost non-intrusive FBT systems for XR \cite{capece2018low}. Some projects look at HPE for XR \cite{li2021hybrik}, as a possible solution to achieve non-intrusiveness.

In addition to not wearing an external tracking marker \cite{desurmont2006nonintrusive}, non-intrusive tracking implies that the tracking system also does not impede the view of the user. This way the user can focus entirely on the experience \cite{park2006non}. When examining the use of HPE for markerless 3D pose estimation, Rahman found that ``3D monocular pose estimation algorithms are promising non-intrusive low-cost markerless methods for evaluating human kinematics'' \cite{rahman2023towards}. 
Recently, some research projects have shown that HPEs can be adapted for differently-abled people, specifically wheel chair users \cite{li2024wheelposer, huang2024wheelpose}. While, these projects use wearable devices, they open possibilities for markerless HPEs be used for non-intrusive FBT and accessibility for everyone. 

We identified the following conditions based on prior work that need to be met to achieve non-intrusive FBT for XR:
\begin{itemize}
    \item C1: User is not required to wear additional hardware.
    \item C2: User is not aware of the tracking hardware. 
\end{itemize}
As a soft requirement low-cost or minimal hardware should also be met. Low-cost FBT aims to allow accessibility to FBT for everyone without large financial challenges. HPEs can be used with RGB cameras \cite{reimer2023evaluation} and in some cases one sensor is all that is required \cite{nie2017monocular}.

C1 can be met through the use of markerless HPE either through depth cameras \cite{capece2018low} or RGB cameras \cite{reimer2023evaluation} for FBT in XR. Further this condition can be fulfilled through out-of-the box FBT from the headsets themselves, e.g. the Meta Quest 3, but this has not been evaluated for differently-abled people yet. 
 
 C2 can be a bit more of a challenge, as this condition of user awareness hasn't yet been evaluated in terms of FBT at the time of this writing. In the context of FBT the tracking camera setup wouldn't surround the user \cite{gabel2012full}, where they are acutely aware of the tracking system. A non-intrusive system would either be integrated into the headset themselves, or the sensor/the sensors should be outside the user's field of view or hidden, this way they can focus on the experience  \cite{desurmont2006nonintrusive}. The tracking system be setup in such a way that users can focus on the experience and not on the tracking system. This is also an advantage for accessibility, where users who may be easily distracted or have difficulty with new situations can focus on the XR experience and not the tracking system.  

\section{Discussion}
With a non-intrusive FBT system accessibility for the setup and experience can be possible \cite{dudley2023inclusive}. Dudley et al. argue that there are seven steps for the setup of a VR system: (i) setting up a VR system; (ii) putting on/taking of the HMD; (iii) adjusting the HMD; (iv) cord management; (v) manipulating dual controllers; (vi) inaccessible controller buttons; and (vii) maintaining controllers in tracking volume. Modern HMDs have greatly reduced the setup complexity for VR systems and issues with cords by offering stand-alone XR experiences. A non-intrusive FBT system would also not raise the complexity as the users only have to interact with their own movements and not focus on controllers. 

Designing 3D interfaces, which are inherently spatial, for interactions for people with motor limitations or people with tremors requires particular attentiveness and considerations \cite{mott2019accessible}.
For example, adapting HPE to counter-balance the individual tremors can readily allow people with tremors to benefit from non-intrusive FBT. Similarly, HPEs can be dynamically adapted for users with upper- and lower-body motor limitations \cite{huang2024wheelpose}. This highlights the potentials and improved accessibility of XR by means of non-intrusive FBT systems. 

\section{Limitations}
This proposal based on our research and experience. Our goal is to spark a discussion between experts in the field of XR and HCI to gather more insight into non-intrusive FBT for accessibility and the possible advantages. Based on this proposal, it is vital for the next step to include an evaluation and identify advantages of the system. The dimensions for this evaluation should encompass user experience, usability and workload. While there are project that examined HPE adapting to users with motor limitations \cite{huang2024wheelpose} the majority of HPE models are trained with able-bodied people. While this might be a limitation for now, we see a need expanding HPE models for all people as a future step.  

\section{Conclusion}
Throughout this paper, it became evident that there is much potential for non-intrusive FBT for meaningful interactions in XR using HPE. Currently, there is only some research that assesses the advantages of HPE in terms of usability and almost none in terms of accessibility.  
This paper proposes conditions for non-intrusive FBT for XR and identifies HPE as the method to fulfill the concept. Next steps include evaluating non-intrusive FBT for XR through expert interviews. Following, this a framework will be developed and implement and then evaluate a non-intrusive FBT system for XR. We are confident that HPE can help provide non-intrusive FBT and bring natural interactions in XR to a broader audience. 



\bibliographystyle{ACM-Reference-Format}
\bibliography{main}

\appendix

\end{document}